\documentclass[conference]{IEEEtran}

\usepackage[utf8]{inputenc}
\usepackage[T1]{fontenc}
\usepackage{microtype}

\usepackage{booktabs}
\PassOptionsToPackage{hyphens}{url}\usepackage{hyperref}

\usepackage{csquotes}

\begin{document}

\title{Requirements Engineering for Research Software: \newline A Vision}

\author{
\IEEEauthorblockN{Adrian Bajraktari}
\IEEEauthorblockA{University of Cologne\\
Cologne, Germany\\
bajraktari@cs.uni-koeln.de}
\and
\IEEEauthorblockN{Michelle Binder}
\IEEEauthorblockA{University of Cologne\\
Cologne, Germany\\
mbinder3@uni-koeln.de}
\and
\IEEEauthorblockN{Andreas Vogelsang}
\IEEEauthorblockA{University of Cologne\\
Cologne, Germany\\
vogelsang@cs.uni-koeln.de}
}

\maketitle

\begin{abstract}
Modern science is relying on software more than ever. The behavior and outcomes of this software shape the scientific and public discourse on important topics like climate change, economic growth, or the spread of infections. 
Most researchers creating software for scientific purposes are not trained in Software Engineering. As a consequence, research software is often developed ad hoc without following stringent processes.    
With this paper, we want to characterize research software as a new application domain that needs attention from the Requirements Engineering community. 
We conducted an exploratory study based on 8 interviews with 12 researchers who develop software. We describe how researchers elicit, document, and analyze requirements for research software and what processes they follow. From this, we derive specific challenges and describe a vision of Requirements Engineering for research software. 
\end{abstract}


\section{Introduction}


Research software plays a pivotal role across diverse disciplines, from modeling climate change~\cite{giorgi_thirty_2019} and analyzing economic markets~\cite{reinhart_growth_2010} to predicting infectious disease spread~\cite{kraemer_effect_2020}. Its influence on scientific discovery is profound, yet it also poses significant risks if results are incorrect or misinterpreted. The case of Reinhardt and Rogoff's paper~\cite{reinhart_growth_2010}, where replication efforts uncovered code errors leading to different conclusions, underscores the importance of rigor in software development for research~\cite{herndon_does_2013}.


The field of Research Software Engineering (RSE) has emerged in response to the critical role of software in research and the specialized skills required for its development~\cite{felderer_toward_2023}. RSE aims to advance research by ensuring software is reliable, efficient, and user-friendly, evidenced by initiatives to provide educational resources and training for researchers.



However, Requirements Engineering (RE) within RSE often remains abstract~\cite{scheliga_dealing_2019}, limited to documentation\footnote{\url{https://www.software.ac.uk/blog/what-are-best-practices-research-software-documentation}}, or overlooked entirely\footnote{\url{https://software-carpentry.org/lessons/}} in educational materials. This lack of specificity fails to address the unique context of research software, leading to perceptions of RE as not applicable or reasonable for research software because \enquote{requirements are not known up-front}~\cite{johanson_software_2018,wilson_best_2014}, \enquote{overly formal software processes restrict research}~\cite{johanson_software_2018}, or \enquote{scientific programmers usually don’t get their requirements from customers}\cite{wilson_best_2014,segal_developing_2008}.

This paper presents findings from interviews with 12 researchers, aiming to understand RE practices and challenges in research software projects. Our investigation reveals a lack of RE best practices, with requirements seldom elicited, documented, or analyzed systematically. Stakeholders are often undefined, and development is characterized by ad-hoc, unplanned approaches, raising doubts about the feasibility or relevance of RE activities.

We propose a vision for ``RE for research software'' articulating necessary adaptations to meet the specific challenges faced in this domain. This vision encompasses both educational aspects and research needs, highlighting a significant gap between scientific questions and explicit software requirements.

Aligned with the conference theme ``Exploring New Horizons: Expanding the Frontiers of Requirements Engineering'', we aim to spotlight research software as an untapped domain for RE application. With the growing significance of research software and its limited interaction with RE, our vision seeks to bridge educational and research gaps, fostering greater maturity in research software engineering and empowering researchers to develop superior software.

\section{Background and Related Work} \label{sec:bg}

\subsection{Research Software and its Engineering}
Felderer et al.~\cite{felderer_toward_2023} define research software as software designed and developed to support research activities in various fields such as science and humanities. It can be used to collect, process, analyze, and visualize data, as well as to model complex phenomena and run simulations. Research software can be developed by researchers themselves or by software developers working closely with researchers. Research software is typically developed to meet specific research needs, and it often has unique requirements that are different from standard commercial software. 
Research software mainly deals with modeling, simulation, and data analytics of, e.g., physical, chemical, social, or biological processes in spatiotemporal contexts. The field of \textit{computational sciences} is a division of science that uses advanced computing capabilities to understand and solve complex physical problems. 

For decades, software engineering and computational science coexisted without much attention to each other. The field \textit{research software engineering} (RSE) emerged in the late 2000s when software development in science, as it was practiced for decades, reached its limits and researchers thus needed to make bigger adjustments to their software for the first time. Through this, a major lack of modern software engineering practices and deficiencies in research software became apparent. This not only influenced the development performance of researchers (referred to as \textit{productivity crisis}~\cite{johanson_software_2018,faulk_scientific_2009}) but also raised concerns regarding the credibility of scientific results (referred to as \textit{credibility crisis}~\cite{hatton_how_1994}).

The following characteristics are often reported as specific to research software~\cite{johanson_software_2018,heaton_claims_2015,basili_understanding_2008}:

\textbf{Few scientists are trained in software engineering:} Researchers are often experts in their scientific domain but were never trained in software engineering. They learn to develop software by talking to senior colleagues (who have also not been trained in SE) and learning while coding.   

\textbf{Unplanned and organic software growth:}
The majority of research software is initially designed for a relatively small scope (e.g., one paper or PhD thesis).    
Only when the software package finds success in the community does it begin to grow. As a result, later modifications become increasingly difficult and error-prone. 

\textbf{Narrow user base:}
Most scientific software is used by its developer or members of the developer's research group. This internal use
leads developers to discount usability (because they can just fix problems as they arise during use), which in turn reduces overall maintainability.

It is important to understand that there is not one monolithic community of scientific software developers. According to Basili~et~al.~\cite{basili_understanding_2008}, three primary variables help developers to better understand how to best integrate software engineering practices into their specific project: team size, envisioned lifetime of the software (one-time use vs.\ community code), and intended users (internal, external, or both).


\subsection{International Survey on Research Software Engineering}
The UK Software Sustainability Institute (SSI) conducted an annual international survey on RSE~\cite{Hettrick2022} since 2016. The survey provides insights into demographics, job details (contract, satisfaction), and practices (coding, tools, publication) applied in RSE. The latest survey from 2022 had 1324 participants from 44 countries, most residing in Germany, the UK, and the US. Most participants have a university level of education, and even 40\% have a doctorate. This correlates with the fact that 62\% stated that they work in a university and 14\% in a national lab. Only 6\% stated that they work in a private company. The top 10 most relevant fields were computer science (43\%), biological sciences (27\%), physics/astronomy (25\%), geography/environmental sciences (16\%), mathematics (15\%), medicine (11\%), chemistry (10\%), mechanical engineering (7\%), education (7\%), and materials technology (6\%). 58\% said they are professional developers, most with between 1 and 15 years of experience. When asked about how time is spent and how they would like to spend their time, around half of the respondents said that more than 50\% of their time is spent developing software, but 30\% indicated they wanted to spend more time on software development. In second place, around 30\% said they spend more than 50\% of their time doing research, but \~33\% want to spend more time doing research. 46\% of the participants said their code is rather used by others than by themselves, and that their peers regularly change. Almost half of the people do not work in a dedicated group. They work on 1--3 projects with 1--3 developers per project. They receive no or very few regular training. Concerning publication, 54\% stated that they are named as co-authors, while others are acknowledged only (24\%) or not mentioned at all (16\%). Half the people always use open-source software, and another 23\% use it very often. Research software is very dependent on each developer, as \textit{bus factors} of 1.0 (53\%) or 2.0 (26\%) indicates. In 83\% of all cases, there is no transition plan for when a developer leaves. RSEs use git (82\%) and testing (on their own (60\%) or through CI (50\%)). The three main programming languages are Python (71\%), C++ (30\%), and R (27\%). In general, RSEs are satisfied with their job but also indicate that recognition outside their group and possibilities to gain a promotion are limited. Most participants are active in an RSE organization.

\subsection{Requirements Engineering for Research Software}

Heaton~et~al.~\cite{heaton_claims_2015} conducted a systematic literature review to identify claims about the use of software engineering practices in science. Related to requirements, they found the following claims:

\textbf{Scientific software developers often do not produce proper requirements specifications:}
Numerous studies have found that scientific software developers do not produce requirements documents~\cite{segal_when_2005,segal_challenges_2009,sanders_dealing_2008,li_domain_2011}. One of the main reasons stated by researchers is that \enquote{full up-front requirement specifications are impossible: requirements emerge as the software and the understanding of the domain progress.}\cite{segal_developing_2008,segal_when_2005,sanders_dealing_2008,johanson_software_2018}

\textbf{When scientific software developers produce requirements, they generally focus on high-level requirements}
High-level requirements are part of the scientist's domain knowledge. They tend to assume that the task of translating these high-level functional requirements into lower-level requirements would be trivial for software engineers~\cite{segal_challenges_2009,li_domain_2011}.

\textbf{When scientists produce high-level requirements, they rely on developers to prioritize them}
The lack of scientific background made it difficult for the software developers to properly prioritize requirements and effectively develop the software. To rectify this problem, a scientific representative had to be assigned at a later stage to prioritize requirements for the developers~\cite{segal_software_2009}.

Easterbrook and Johns~\cite{easterbrook2009} conducted a case study of climate scientists' software development at the UK Meteorological Office’s Hadley Centre for Climate Prediction and Research, a government-funded research lab. We will briefly summarize the main observations.
Software development at the lab was organized similarly to the onion model from open-source projects. At the core, 12 developers from IT support are responsible for accepting changes to the core model. At the next level, 20 senior researchers act as code owners of selected parts of the core model. The outer layer consists of researchers using the model in their research field. Releases were done regularly, around every 5 months. For the planning of a release, agile planning techniques were used. Every change to the core model must pass two review stages, first by the relevant code owner for a scientific check, then an IT member performs a system review. The scientists maintained a shared understanding of their software mainly via face-to-face communication. Often, cross-functional teams are formed to investigate specific issues. A wiki provides to-dos, design notes, status reports, glossaries, etc. Any code-related representations, if used at all, are either flow charts or design descriptions tackling the underlying equations. The models are understood as imperfect representations of complex physical phenomena, thus they do not reason about code errors but view them as evolving theories, solely serving to test hypotheses.

Although such empirical investigations about RE activities in research software have already been conducted between 2005 and 2010, we found only a few papers where authors suggested adapted RE practices for research software.
Li~et~al.~\cite{li_effective_2015} present DRUMS Board, which supports lightweight
creation and management of requirements in scientific contexts.
In a controlled experiment in the domain of computational fluid dynamics, the authors found that using DRUMS Board has a significant effect on the number and innovativeness of generated ideas in the requirements elicitation process. In a follow-up paper, Li~et~al.\ suggested another approach to support scientists in gathering requirements by automatically extracting them from existing project reports and manuals~\cite{li_automated_2015}. 
Already in 2006, Smith proposed a customized requirements specification (SRS) template for research software~\cite{smith_systematic_2006}.

The German Aerospace Center (DLR) published a guideline on how research software should be developed at DLR~\cite{schlauch_dlr_2018}. The guideline divides research software into four application classes 0--3. These resemble criticality levels ranging from small scripts for personal use (class~0) over small-scoped software that should be understandable to other DLR researchers (class~1), long-term development like larger frameworks (class~2) to mission-critical software, software with warranty, or software contributing to funding (class~3). Each class adds a set of recommended practices and constraints to those from the lower classes. For requirements engineering, the guideline states that basic RE practices should be applied beginning from class~1, e.g., collaborative problem definition and documentation. In class~2, the guideline recommends employing documentation and prioritization of functional and quality requirements, documentation of user groups and tasks, a glossary, and continuous refinement of a list of requirements. For class~3, the guideline recommends traceability for all requirements, guidelines for documentation of requirements, and active risk management.

\subsection{RE for Related Domains}
Research software shares some characteristics with other domains for which RE practices have been investigated in the past.

\textbf{RE in ML and Data Science Projects}:
Some research software implements a data science or machine learning approach to answer a scientific question. Requirements Engineering for data-intensive and ML systems has also been reported to be challenging~\cite{Vogelsang2019}. Ishikawa and Yoshioka~\cite{ishikawa_how_2019} found that requirements engineering is the most difficult activity for the development of ML-based systems, as it is impossible to estimate or assure the system's accuracy in advance.
Villamizar~et~al.~\cite{villamizar_requirements_2021} found several challenges of RE for ML, including the lack of validated techniques specific to ML, e.g., ML-specific requirements specification, and a knowledge gap in how to deal with ML-specific quality requirements.

\textbf{RE in Open Source Software}:
Research software is often developed open source. Open source software often relies on contributions from different contributors who do not belong to a single team. Requirements are elicited by a core development team from contributions through change requests, analyzed and prioritized through automation and community feedback, and documented informally through decisions and voting~\cite{Tasnim2023}.

\textbf{RE in Agile Development}:
The main goal of agile development methods is to deal with changing requirements that are unknown up-front. Thus, it may seem like a perfect fit for developing research software. 
Sch{\"o}n~et~al.~\cite{Schoen2017} overview the literature on agile RE practices, which may also fit the research context. However, as we will later see in our results, researchers rarely apply agile practices. 


\section{Study Design} \label{sec:method}
Our goal was to understand requirements-related practices in RSE in more detail to find out why they are rarely applied and what needs to change to make RE a beneficial activity for developing research software. 
We were interested in the following questions:
\begin{itemize}
\item\textbf{RQ1:} How do researchers elicit, document, and analyze requirements for research software?
\item\textbf{RQ2:} What processes do researchers follow and which parts relate to requirements?
\item\textbf{RQ3:} What are requirements-related challenges that researchers face?
\end{itemize}

\subsection{Research Method}
We employed a qualitative approach, using semi-structured interviews. Semi-structured interviews employ an interview guide with questions but allow the order of questions to vary to fit the natural flow of the conversation.
Our interview questions took an exploratory approach. We asked respondents about their background and the practices in a recent research software project they were involved in.
Afterward, we specifically asked for ways how they elicit, communicate, document, and test expectations and requirements for their research software. The interview guideline is available in the supplementary material\footnote{\url{https://doi.org/10.6084/m9.figshare.25197161}}


\subsection{Study Subjects}
Since research software is often connected to relatively complex theories (e.g., simulations of complex mathematical models, complex data analytics), most research software is driven by people who are experts in the application domain (e.g., physicists, meteorologists, economists). 
They are the people who currently prepare the data, make design decisions, and finally evaluate the performance of their systems. 
Therefore, we selected these researchers as subjects.
Table~\ref{tab:participants} gives an overview of our participants, their scientific field, and years of experience. In total, 12 scientists from various scientific disciplines were interviewed in 8 interviews, with each interview lasting approximately one hour. For confidentiality, respondents are kept anonymous and referred to with running IDs P1--P8. 
The first five interviews were conducted by all authors, the remaining three interviews were conducted by only one of the authors. All interviews were recorded and transcribed. 
\begin{table*}
    \centering
    \caption{Study participants}
    \label{tab:participants}
    \begin{tabular}{@{}lrlllll@{}}
    \toprule
    \textbf{ID}& \textbf{\# part.\ }&\textbf{Scientific field}&\textbf{Experience}&\textbf{Time invest in SW}&\textbf{\# devs}&\textbf{internal/external?}\\
    \midrule
        P1 & 2 & Mathematics (Numeric) & 25+y & many hours/week & 16--17 & internal\\
        P2 & 1 & Earth System Data Exploration & 8y & "much" & 1-2 & internal\\
        P3 & 1 & Geophysics and Meteorology & 26y & none currently & 5 & internal\\
        P4 & 1 & Geography & 2y & 1--4 hours per day & 1 & internal\\
        P5 & 3 & Biology/Bioinformatics & 10--35y & 80--100\% of time & 3 teams & external\\
        P6 & 1 & Aerospace Engineering & 13y & 50\% of time & 2 & internal\\
        P7 & 1 & Astrophysics & 15y & 90\% of time & 1--5 & both\\
        P8 & 2 & Theoretical Computer Science & 3+y & few to many hours & 2 & internal\\
    \bottomrule
    \end{tabular}
\end{table*}

\subsection{Data Analysis} \label{sec:analysis}
We relied on thematic coding~\cite{Gibbs08} in a collaborative setting. Each interview was coded by two authors independently. The codes, results, and lessons learned were discussed, validated, and merged in two meetings with all three authors. Statements in the transcripts were assigned one or more codes. 
In particular, we used a mixture of a priori coding based on our research questions, and emergent coding starting from any mention of requirements. 
We then combined data from all transcripts in a meeting to ensure that we covered the full data in our synthesis of findings. 
Quotes have been translated from the respondents' native language to English and edited for readability. Colloquialisms have been kept to convey the tone of the conversation and to reflect the informal nature of the interview setting.
The resulting codebook is available in the supplementary material\footnotemark[3].

\subsection{Threats to Validity}
Maxwell~\cite{Maxwell12} identified five threats to validity in qualitative research that also apply to our study design. \emph{Descriptive validity} refers to the threat that an interviewer does not collect all relevant data during an interview. To mitigate this threat, we recorded the interviews on tape. We annotated the transcripts with pointers to the respective positions in the recording to trace back to the original conversation. \emph{Interpretation validity} refers to the possibility of misunderstandings between interviewees and the researchers. The study goal was explained to the participants before the interview to minimize this risk. Steps taken to improve the reliability of the interview guide included a review. \emph{Researcher bias} and \emph{theory validity} are threats that refer to the researcher's bias to interpret the interviews in a way that serves his or her goals or initial theory. Since this is our first study in this area, we do not have a specific RE methodology that we would like to promote. That means we were very much open to the outcomes of the interviews. Furthermore, we tried to lower researcher bias by cross-validating the annotated codes by two authors. \emph{Reactivity} refers to the threat that interviewees behave differently because of the presence of the interviewer. Getting rid of reactivity is not possible, however, we are aware of it and the way it may influence what is being observed.

\section{RE Practices in Research Software Engineering} \label{sec:reqts}
In this section, we answer RQ1. For this, we give an overview of the most important requirements mentioned by the interviewees. Then, we will report on the practices applied for \textit{elicitation}, \textit{analysis, validation and verification}, \textit{documentation}, and \textit{management}.

\subsection{Most Important Quality Requirements}
Our interviewees highlighted several key quality attributes essential for their research software. 
Above all, functional correctness is vital, as the software would otherwise have no use for them. Thus, the focus is put on functional correctness first. 
Following functional correctness, they prioritized performance (P1, P3, P4, P7, P8), maintainability (P5, P6), portability (P1, P2), and installability (P1, P2), which were frequently mentioned. Other attributes such as scalability (P2), interoperability (P1), extensibility (P1), understandability (P1), and reusability (P5) were mentioned once, suggesting they might be project-specific. However, nearly all participants (except P3) emphasized reproducibility as a crucial quality; P5 noted, \enquote{Reproducibility, for example, I think it's one of the most important points in our research.} This emphasis on reproducibility distinguishes research software from commercial/non-research software, underscoring its fundamental role in research.

Although all interviewees recognized reproducibility as crucial, they struggled to describe specific practices supporting it. Only the use of Docker (P5, P7) and the Julia programming language (P1) were mentioned, with P1 asserting that their software ensures complete reproducibility of published papers within three to five minutes. Challenges to reproducibility were also highlighted, including the need for early planning (P2), difficulties due to large data volumes or high-performance hardware requirements (P2), and the challenge of reproducing environments in tools like Jupyter notebooks (P2, P7). Overall, it appears that our interviewees lacked clear strategies for actively ensuring reproducibility.


\subsection{Elicitation}
Our interviews indicate that researchers often skip extensive requirements elicitation. However, a few interviewees did mention engaging in some form of requirement gathering, whether as an isolated occurrence or infrequently. For instance, one used expert interviews during their master's thesis (P4), while others experimented with prototyping (P4, P5) or user demonstrations (P6). One mentioned relying solely on focus group meetings to refine a particular standard that guides their software's development (P6), and another team bases their requirements solely on project plans or scientific concepts (P1). Conversely, some reported lacking any user feedback or structured requirements elicitation process (P1) or mentioned that requirements develop spontaneously due to the varied input from multiple contributors (P2).


\subsection{Documentation}
Documentation practices among our interviewees show significant variation, falling into three distinct levels of granularity. The first level includes those who forgo documentation entirely, prioritizing feature development and scientific advancement over documentation due to time constraints (P3). The second level consists of participants who document but with minimal structure, such as using to-do lists in code comments (P8) or maintaining Word documents (P1, P8). The third level encompasses those who adopt a more structured approach to documentation, using tools like issues and pull requests (P1, P2, P6, P7), Kanban boards (P2), change logs (P2, P4), Wikis (P7), or automatic documentation generation (P2, P4). Even among those employing more sophisticated documentation methods, it is considered a lower priority and often neglected due to time constraints (P2, P6, P7).

\subsection{Analysis, validation, and verification}
Researchers often do not systematically elicit requirements, frequently encountering new needs during the coding process. Similarly, documentation of requirements, if done, lacks structure. As a result, conducting any form of requirements analysis, such as resolving conflicts or ensuring consistency, alongside formal validation and verification (V\&V), proves difficult. None of our interviewees reported engaging in detailed analysis or V\&V activities beyond performing informal checks to see if the software broadly functions as intended. This often involves running simulations and evaluating the outcomes to determine their sensibility, as described by P3: \enquote{We let simulations run and then look at the results. We check whether the results make sense, and if not, then there must be a bug.}

\subsection{Requirements Management}
Requirements management shows clear shortcomings across our interviews. The majority of participants acknowledge a lack of a structured approach to managing requirements. For example, P1 noted, \enquote{As of now, there are no formal requirements or infrastructure requirements that we somehow manage together,} indicating the absence of a centralized requirements repository, such as a product backlog (P1, P5). P6 mentioned their requirements are mostly outlined in a standard, suggesting little need for additional requirements management. P2 described a system of prioritizing work necessary for the next milestone, though this priority list is subject to frequent changes due to scientific developments (P2, P3).
Furthermore, it was observed that almost none of the participants developed test cases directly from requirements (P1, P7, P8). Instead, test cases are often based on personal experience (P3, P4, P5). In data-centric environments, the challenge of deriving test cases from data was highlighted, mainly because researchers are unsure how to undertake this process (P5).

\section{RE Processes in Research Software}
In this section, we answer RQ2. 
For this, we will cover the \textit{process model}, \textit{meetings}, \textit{software evolution}, and \textit{workflows}.

\subsection{Process Model} 
Our interviews revealed that many researchers work in small teams of 2 to 3 people (P2, P5, P8), or even individually (P4). Given the limited team size, they view structured software development processes, such as Scrum, as unnecessary overhead without perceived benefits, especially for solo developers. Researchers are often uncertain about the future significance of software developed for specific papers, leading to reluctance to invest in quality assurance. Previous attempts to follow Scrum were met with criticism, particularly regarding its fixed-time iterations, which participants found incompatible with research goals (P5). The unpredictable nature of research objectives makes it challenging to plan sprints, as the exact tasks required to achieve a goal are not always clear beforehand.

\subsection{Meetings}
Our findings indicate that discussions about software requirements are rare among researchers. The majority of interviewees do not participate in regular meetings specifically dedicated to software discussions (P3, P7, P8), with only two exceptions who hold such meetings weekly (P1) or monthly (P6), respectively. P8 mentioned, \enquote{[Most communication] between us is in passing. Sometimes there is a todo in code,} highlighting an informal and sporadic nature of communication about software. Researchers prefer dedicating meeting time to discuss scientific problems, viewing the software primarily as a tool to achieve their research objectives. This lack of focused dialogue on software requirements oftentimes results in oversight or neglect of these requirements, as the primary emphasis is on scientific inquiry rather than on the software development process itself.

\subsection{Software Evolution}
Our interviewees highlighted a common challenge: the lack of predefined requirements for their software. They pointed out the absence of a clear vision or planned set of features, encapsulated by P1's observation: \enquote{We only add to it [the software]. We never finish something, which is why the software is growing. And it's growing strangely. It's uncontrolled.} This reflects a pattern where requirements are identified and shaped during the development process (P7), leading to an unplanned, organic, and opportunistic expansion of the software (P1, P2, P4).

\subsection{Change and Release Workflows}
The majority of our interviewees do not adhere to a stringent development workflow, with only two participants using the \textit{feature-branch} workflow for software development (P2, P7). Furthermore, most do not implement fixed release cycles (P3, P4, P7, P8) and, as a result, lack defined milestones (P5) and formal release management practices (P1, P4). Instead, they often consider the current state of the git repository as the ``official current release.''
This approach is linked to the absence of a concise project vision and the challenges in predicting the future functionalities of the software. P1 explained: \enquote{We do not have a defined goal. We don't write our software like this: We have this project and in three years, we achieve X. And by then, the software must be finished, with this catalog of features.} This statement underscores the dynamic and unpredictable nature of developing research software, where long-term planning and feature forecasting can be exceptionally challenging.

\section{RE-related Challenges for Research Software} \label{sec:re-challenges}
In this section, we report the challenges hindering researchers from applying RE practices (RQ3).

\subsection{Challenge 1: Deriving Requirements from Research Ideas}
Our interviews revealed that transforming research ideas into concrete software requirements is notably challenging, differing significantly from the process in production software environments. Researchers and stakeholders often lack foresight into specific requirements (P4), beyond the overarching research questions they aim to explore. A distinct vision for the software is commonly absent (P1, P2, P4), and there is a lack of effective strategies or tools to decompose research questions into actionable requirements (P1, P5). Typically, coding begins with requirements being identified as development progresses (P7). This points to a fundamental challenge: finding effective ways for researchers to systematically translate scientific questions and objectives into software requirements.

\subsection{Challenge 2: Unclear Benefit of SE Practices}
The interviews highlighted a prevalent lack of awareness regarding Software Engineering (SE) practices among researchers. There is a prominent emphasis from management, colleagues, and supervisors on prioritizing research goals, writing and publishing papers, and developing features that support these objectives, rather than focusing on enhancing code quality or adopting structured development processes (P1, P5, P6).
This deficiency in awareness is not limited to those in supervisory roles; the researchers themselves are often unfamiliar with, or even resistant to, SE practices. This ranges from not knowing specific terminology, such as ``Requirements Engineering'', to outright dismissing the value of SE practices based on misconceptions or misunderstandings. The common belief is that implementing SE practices would not significantly benefit their development efforts, especially considering their software is typically developed with a short-term focus (P7). Researchers argue that the time and effort required to apply SE practices do not provide immediate rewards but add unnecessary complexity to their workload.
However, it is notable that many research software projects begin as small-scale initiatives, such as a PhD project or a tool for a single paper, before evolving into significant research and teaching resources. This evolution underscores the potential long-term value of integrating SE practices from the outset, despite the initial perception of them as mere overhead.


\subsection{Challenge 3: Lack of RE Knowledge}
A unanimous point among our interviewees was the absence of formal SE education in their backgrounds. While a few had undergone structured programming courses, the majority acquired their SE and coding skills through self-study, tailored to immediate needs (P4, P7). This ad-hoc learning approach became evident when some interviewees were unfamiliar with the term ``Requirements Engineering.''
Several interviewees expressed a desire to learn SE best practices but found themselves overwhelmed by the sheer volume of available practices, tools, and processes. The daunting task of comprehensively understanding these elements led to early discontinuation of their efforts to formalize their SE knowledge.
This knowledge gap in programming and SE not only results in poor code and process quality but also significantly extends the onboarding period for new team members. Newcomers must simultaneously learn the necessary technologies and practices while understanding the current software's workings and structure. Some participants noted that the onboarding process could last months (P1) or even up to three years (P7) before a new researcher could make meaningful contributions.
%

\subsection{Challenge 4: No SW Development Process}
Almost all participants indicated they do not adhere to any formal software development process. However, their described methods bear resemblances to agile practices, prompting the question: why not implement agile methods more formally? The reasons provided include the perception of agile methods as an unnecessary overhead for small teams (P2, P5, P8), with no perceived added value. Moreover, the use of fixed-length iterations, a core aspect of frameworks such as Scrum, clashes with the exploratory nature of research software development (P5)---particularly since many projects begin with the modest scope of supporting a single publication, making the introduction of comprehensive processes seem excessive initially.
The interviews, corroborated by prior studies, suggest that the scenario wherein researchers develop software independently often results in software of inferior quality. Equally, outsourcing development to a dedicated team without proper integration with the researchers can be ineffective. This is highlighted by P5's observation: \enquote{I think the researchers are not aware of these things, so they regard the bio-informatics facility as a black box. You send them the data and they get the results. It's important for them, but you should make them aware of this that it is important to work on this.}
This underscores the need for a process that accommodates the unique characteristics of research software development---balancing the flexibility required for exploratory research with the benefits of structured agile methodologies, thereby enhancing software quality without imposing undue burden on small developer teams.

\subsection{Challenge 5: How to Achieve Reproducibility?}
Reproducibility emerged as a key concern among nearly all interviewees. Despite acknowledging its importance, they struggled to articulate specific strategies for ensuring reproducibility, other than mentioning some tools believed to address this need. Their inability to define reproducibility's scope highlights a broader issue: while reproducibility is critical for research software---contrasting with its lesser significance in production software---it lacks a clear, universally accepted definition. The ambiguity extends to understanding what aspects are unique to reproducibility versus those that overlap with other software quality attributes.

\section{RE for Research Software: A Vision} \label{sec:vision}

Based on the characteristics and challenges of research software and its context, we claim that more research is needed to address the specifics of research software while customizing the RE process. In the following, we outline our vision for RE for RS and identify research gaps that need to be filled. 

We envision that RE for RS should support agile and exploratory development. In terms of the three process facets defined in the IREB CPRE FL handbook~\cite{IREB-handbook}, most research software is iterative (\textit{time}) and exploratory (\textit{purpose}). The \textit{target} facet is unclear since most research software starts with a customer-specific target (to answer a particular research question) but may then shift to a market-oriented target when it is published and considered useful by other researchers. The suggested RE process should therefore incorporate continuous interaction between researchers, stakeholders, and developers to create and manage visions, prototypes, and product backlogs. In this process, the following specifics need to be addressed.   

\subsection{Elicitation: From Isolated Researchers to Stakeholder Groups} 
Research software is often developed by researchers alone or in small teams. Our interviews reflect that researchers rarely talk to other people about requirements. Most of the time, an overview of stakeholders for a particular software does not exist, which makes it impossible for researchers to elicit all relevant requirements. The fact that research software is often not reproducible, not reusable, or not interoperable is also a result of missing stakeholder management during the development. 
A list of typical stakeholders and their respective interests may help researchers consider all relevant perspectives and produce software with higher value for all stakeholders. 
Typical stakeholders and interests may include:
\begin{itemize}
    \item \textbf{The researcher}: Needs the software to answer a scientific question efficiently. Main interests: Functional Suitability, Performance
    \item \textbf{Reviewers}: Needs the software to reproduce and verify the results of a scientific contribution. Main interests: Reproducibility, Ease of Use
    \item \textbf{Supervisors\slash Managers}: Need the software to create synergies, attract new researchers to work with them, and maintain the scientific reputation and integrity of their group. Main interests: Reusability, Correctness, Performance, Maintainability
    \item \textbf{Other researchers}: Need the software to use or extend it for their scientific purposes. Main interests: Reusability, Ease of Use, Interoperability 
    \item \textbf{Society}: Need the software to be correct and valid w.r.t.\ a scientific theory to assure the reliability of the contributed scientific knowledge. Main interest: Correctness, Reproducibility, Validity   
\end{itemize}

There may be additional stakeholders in certain project settings such as \textbf{several PIs} involved in a joined project (P1) or dedicated \textbf{research software engineers} helping to implement a scientific question in software (P4, P5). 

\subsection{Elicitation: Reproducibility}
Reproducibility is fundamental to the scientific method, ensuring that the outcomes of a study can be consistently replicated under similar conditions. In computational sciences, reproducibility specifically requires that all data and code be made available, enabling identical computational results upon re-execution.
Despite consensus on its significance, our interviewees struggled to articulate how reproducibility translates into actionable software qualities. This gap underscores the necessity of further examination of reproducibility as a distinct quality requirement. It potentially intersects with other software quality attributes, including portability, installability, analyzability, co-existence, and learnability.
Future studies should delve into the concept of reproducibility in software, distinguishing its unique facets from those covered by existing quality attributes. Identifying these nuances will enhance our understanding and implementation of reproducibility in research software. Clarifying these aspects will not only facilitate the development of more reproducible research software but also help in formulating specific practices and tools to achieve it effectively.

\subsection{Documentation \& Analysis: From Research Ideas to Requirements} 
As mentioned before, our interviewees mentioned a big gap between a scientific question and concrete requirements for their software. RE for RSE needs to provide methods and tools that help researchers break down a specific question into requirements. Already in 2006, Smith proposed a customized requirements specification (SRS) template for research software~\cite{smith_systematic_2006}. In a blog post\footnote{\url{https://bssw.io/blog_posts/user-stories-in-scientific-software-development}}, Marques and Milewicz suggested the use of user stories because they noticed \enquote{greater need for communication and consensus in scientific software development} and \enquote{need for better strategies for planning and prioritizing development work}. When applied in one of their projects, they found that \enquote{most user stories the team created were about process improvement, that is, finding better ways of thinking about and working with scientific software.} and gave the following example: 
\enquote{As an application architect, I want to better understand version control capabilities that allow integration of independently developed components so that we can distribute a coherent software stack.}
Despite this reported positive experience, we think that user stories and epics may also be well-suited to bridge the gap between a high-level research question and concrete requirements for the software. Goal-oriented RE~\cite{Horkoff2017} may be another option to consider for this purpose.

\subsection{Management: From Product Owner to Research Owner} 
One of the distinct aspects of research software lies in its perception by developers; rather than viewing it as a product, researchers consider software primarily as a tool to address scientific problems. Consequently, researchers often assume the product owner role themselves, developing the software chiefly for their use. Supervisors seldom adopt this role due to either a lack of relevant knowledge or a focus on scientific outcomes over the software itself. 
In the context of research software, we suggest reimagining the product owner role as a \textit{research owner}. This role involves upholding the research question's vision while recognizing the software's possibilities and limitations. The research owner is tasked with defining and ordering requirements with stakeholders' interests in mind. Supervisors or senior scientists could aptly fill this position, leveraging their experience and perspective.
Moreover, research owners should facilitate and participate in regular discussions on the software's progress, akin to sprint planning meetings in agile methodologies. This approach ensures continuous alignment between software development and the overarching research objectives, enhancing both the process and the outcomes of scientific inquiry.   

\subsection{Management: From Scrum to Agile Research Processes} 
In Section~\ref{sec:re-challenges}, we discussed how researchers often overlook the necessity of adopting formal SE processes, finding methodologies like Scrum too rigid for their needs. This highlights the requirement for a development model more attuned to the nuances of research software development---one that can adapt and scale with the project's evolution in scope, developer base, and user community.
Many research software projects commence as minor endeavors, perhaps initiated by a single PhD student who may not initially recognize the importance of documenting requirements, holding regular meetings, or ensuring the software's maintainability. However, as these projects gain traction and more researchers express interest in using, extending, or contributing to the software, the absence of structured processes becomes a significant hindrance. This is often realized too late, as captured by P3: \enquote{I've recently seen it twice [\ldots] it's enough for one publication but afterward, the software is not usable at all and you start all over again.}
To address these challenges, we propose a development process that encourages early and continuous collaboration between scientific domain experts and software development specialists. This process should be flexible yet structured enough to introduce essential SE practices without hindering the exploratory nature of research. By bridging the gap between scientific inquiry and systematic software development from the outset, we can create more robust, maintainable, and scalable research software that fulfills its intended scientific objectives while remaining open to extension and collaboration.


\section{Conclusions}
Software is increasingly vital for scientific advancement, akin to its role in many industrial sectors. The emerging field of research software engineering addresses this significance by tailoring methods and tools to the unique requirements of research software, alongside providing educational resources for researchers involved in software development. Our study reveals that Requirements Engineering (RE) practices are largely overlooked in this context, attributed to the perceived lack of benefits for individual developers or small teams, a deficiency in necessary knowledge, or software being developed solely to address specific research questions. We propose a vision for RE in research software, highlighting the need for further study on effective stakeholder management within scientific projects, defining reproducibility as a quality requirement, creating development processes that accommodate exploratory work and organic growth, refining the role of the product owner, and establishing methods to translate research concepts into software requirements. Through this paper, we aim to draw the RE research community's attention to research software as a compelling and impactful area of study.

\section*{Acknowledgment}
We thank the interview participants for their commitment and input. We thank our students for their help with transcribing the interviews. 

\bibliographystyle{IEEEtran}
\bibliography{references,rse-references,re4ml-references}

\end{document}